\documentstyle[prl,aps,twocolumn]{revtex}

\input{epsf.def}

        \def\half{{\textstyle \frac{1}{2}}}
        \def\tr{\hbox{Tr}}

\def\be{\begin{eqnarray}}
\def\ee{\end{eqnarray}}
\def\bee{\begin{eqnarray*}}
\def\eee{\end{eqnarray*}}

\def\shan{{\rm Shan}}

 \def\ts{\textstyle}

\def\rt2{\ts \frac{1}{\sqrt{2}} }

\def\half{{\textstyle \frac{1}{2}}}
\def\frth{{\textstyle \frac{1}{4}}}
\def\amp{ {\rm Amp} }
\def\sqz{ {\rm Sqz} }
\def\str{ {\rm Str} }

\def\bw{{\mathbf w}}
\def\dtsig{\cdot \sigma}

        \title{Qubit Channels Can Require More Than Two Inputs to
Achieve Capacity}

\author{Christopher King\thanks{king@neu.edu} and Michael Nathanson}
\address{ Department of
        Mathematics,  Northeastern University,
 Boston, MA 02115}
        \author   {Mary Beth Ruskai \thanks{ruskai@mediaone.net}}
 \address{ Department of
        Mathematics, University of Massachusetts  Lowell,  Lowell,
        MA  01854 USA}

\begin{document}

\date{\today}

 \maketitle

\begin{abstract}

We give examples of qubit channels that require three
input  states in order to achieve the Holevo capacity.

\end{abstract}

\pacs{ }

The Holevo capacity $C(\Phi)$ of a channel $\Phi$ is defined as the
supremum
over all possible ensembles ${\cal E} = \{\pi_j, \rho_j\}$ 
(consisting of  a probability distribution $\pi_j$ and 
set of density matrices $\rho_j$), of the quantity
\be \label{eq:holv}
\chi( {\cal E}) =   \chi( \{\pi_j, \rho_j\}) = S[\Phi(\rho)] -   \sum_j \pi_j
S[\Phi(\rho_j)]
\ee
where   $\rho = \sum_j \pi_j \rho_j$ is the average input,
and $S(\gamma) = -\tr \gamma \log \gamma$ denotes the von Neumann
entropy.  Thus, $C(\Phi) = \sup_{{\cal E}} \chi( {\cal E})$.
It has been shown \cite{Hv2,SW1} that $C(\Phi)$
\linebreak is the maximum
information carrying capacity of a channel restricted to product inputs,
but permitting entangled measurements.  $C(\Phi) \geq C_{\shan}(\Phi)$, 
where the classical Shannon capacity $C_{\shan}(\Phi)$
 describes the information carrying capacity of a channel
when output measurements (as well as input ensembles) are restricted
to products.  (See, e.g.,  \cite{Fuchs,Hv3,KR2} for  precise definitions.)
We will sometimes  use subscripts to denote the supremum of
(\ref{eq:holv}) restricted to a particular class of ensembles;
in particular, we write
$C_{n}(\Phi)$ to denote the restriction to ensembles of
$n$ states. It is well-known
\cite{Davies} that for qubit channels the maximum can be achieved with an
ensemble containing at most four states.

  \smallskip

In general, a qubit channel maps the Bloch sphere to an ellipsoid
\cite{KR1,RSW}.
If the channel is unital, i.e, if $\Phi(I) = I$, then the
ellipsoid is centered at the origin, and the
capacity is achieved with a pair of orthogonal inputs whose
images, which are also states of minimal entropy \cite{KR1}, are the
endpoints of the major axis of the ellipsoid.  For non-unital channels
the ellipsoid is displaced from the origin, and examples are
known \cite{Fuchs,SW2} for which the capacity
is achieved with two
non-orthogonal inputs which are not mapped onto states
of minimal entropy.   Here we
present examples of non-unital channels which require {\it three} input states
to achieve capacity. Furthermore these channels are {\it non-extreme}
points in the set of
all qubit channels, and this property is essential to
our construction.

To motivate our strategy, we consider the capacity of two
well-known channels which have rotational symmetry about an axis of the
Bloch sphere, and we maximize
(\ref{eq:holv}) over ensembles consisting of a pair of states
on a line either parallel or orthogonal to this axis.

\begin{figure}
\epsfxsize=9cm
\epsfbox{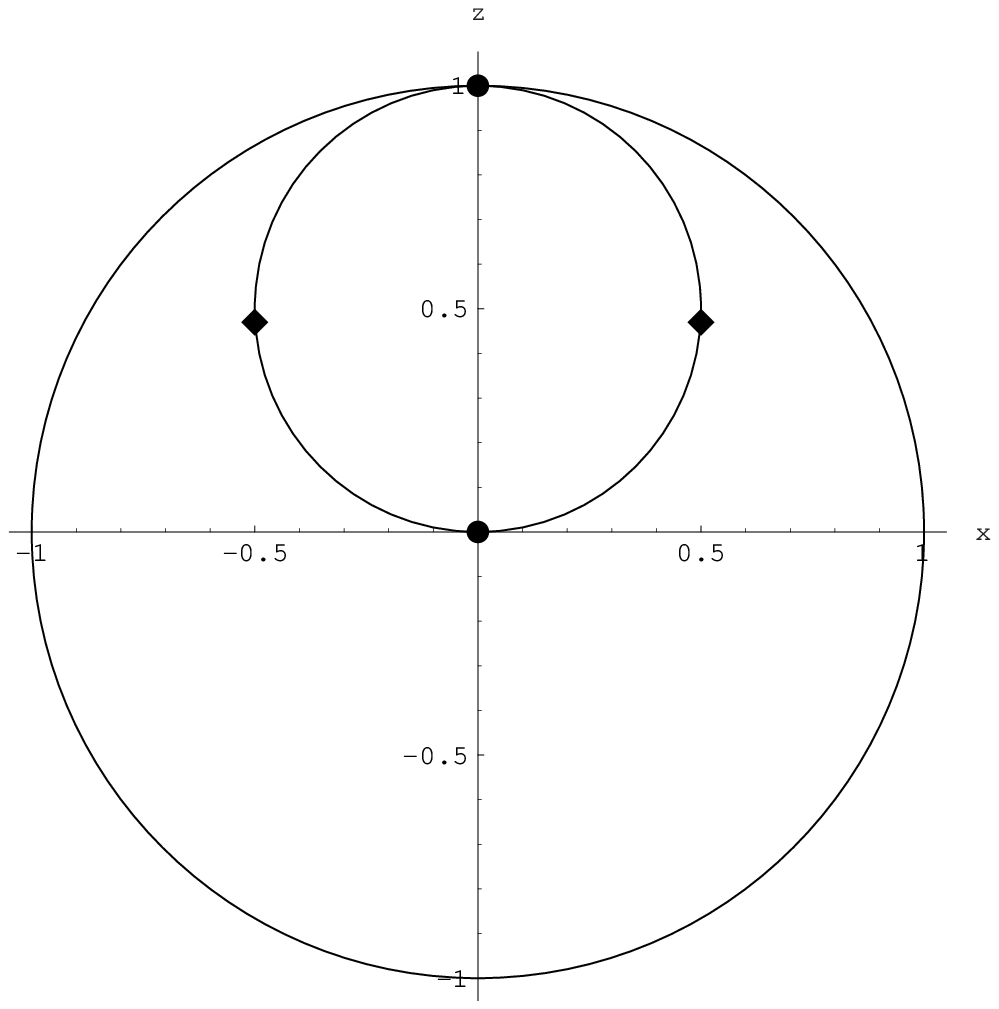}
\caption{The x-z plane of the maximally shifted depolarizing
channel $\Phi_{D}$ with radius $\mu = 0.5$.  The images of  the
 optimal vertical and horizontal ensembles are indicated
by $\bullet$ and $\diamond$ respectively.}
\label{fig:depol}
\end{figure}

First, let $\Phi_D$ denote the shifted depolarizing channel which
contracts the Bloch sphere to a sphere of radius  $\mu$ and shifts it
up until it touches the unit sphere, i.e, if
$\rho$ is written in the form $\rho = \half \big[I + \bw \dtsig \big]$,
then
\be \label{eq:dep}
  \Phi_{D}(\rho) & = & \half
 \big[I + (\mu w_1, \mu w_2, (1-\mu) + \mu w_3) \dtsig \big] .
\ee
When $\mu = 0.5$,  the image of the Bloch sphere  satisfies
\be
   x^2 + y^2 + (z-\half)^2 = \frth
\ee
as shown in Fig.~\ref{fig:depol}, and the poles
$\rho = \half \big[I \pm \sigma_3 \big]$ are mapped to
the  states $\half \big[I + \sigma_3 \big]$ and $\half I$
which have entropy $0$ and $1$ respectively
(using base 2 for logarithms).
If we restrict the input ensemble to a vertical line,
the $\chi$-quantity (\ref{eq:holv}) is maximized by a convex
combination of the poles, so that the average output state
$\Phi_D(\rho) = \half \big[I + z \sigma_3 \big]$ lies on the
$z$-axis. One finds that this vertical capacity,
which we denote $C_V$, is
achieved when the output average is at $z = 0.6$ and its value is
\bee
C_V & = & \sup_{z} \left\{ S\Big( \half \big[I + z \sigma_3
\big] \Big) -  (1-z)  \right\}  \\ & = &
  S\Big( \half \big[I + 0.6 \sigma_3 \big] \Big) - 0.4 \approx 0.32193.
\eee
(This is also the capacity of a 
quantum-classical channel $\Phi_{QC}$, discussed below, which maps
$\rho \mapsto \half \big[I + (\half + \half w_3) \sigma_3 \big]$.)
If, instead, we restrict to \linebreak convex combinations of
states on a horizontal line, the restricted horizontal capacity
$C_H = 0.2144$ is achieved  when the output average is at the
midpoint of the line with
 $z = 0.474$.  Numerical studies confirm that $C_V$
is the unrestricted maximum of  (\ref{eq:holv}).

\begin{figure}
\epsfxsize=9cm
\epsfbox{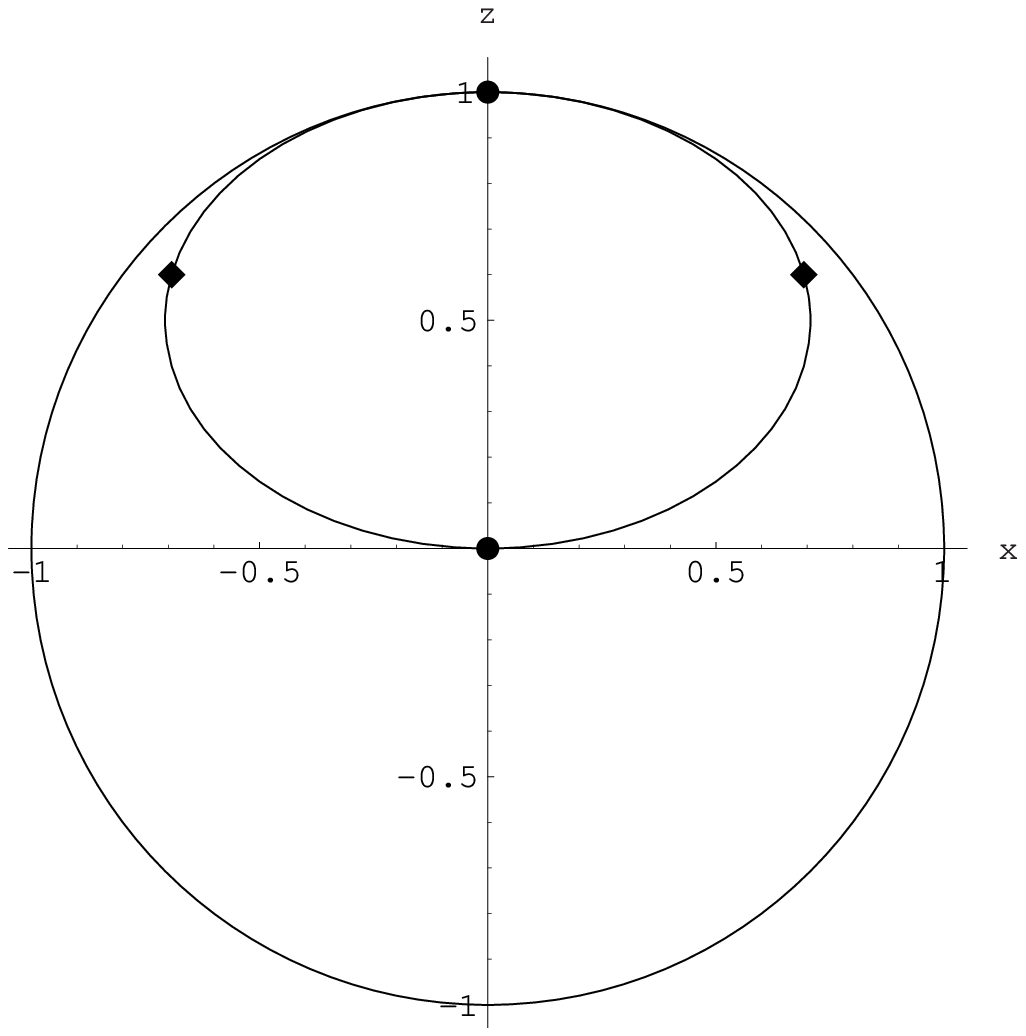}
\caption{The x-z plane of the amplitude damping channel $\Phi_{\amp}$
with $\mu = 0.5$. The images of  the
 optimal vertical and horizontal ensembles are indicated
by $\bullet$ and $\Diamond$ respectively.}
\label{fig:amp}
\end{figure}

Next, let $\Phi_{\amp}$ denote the amplitude damping channel
\be  \label{eq:amp}
\Phi_{\amp}(\rho) & = & \half
 \big[I + (\sqrt{\mu} \, w_1, \sqrt{\mu} \, w_2, \, (1-\mu) + \mu w_3)
\dtsig \big] .
\ee
When $\mu = 0.5$, the poles  are again mapped into the states
$\half \big[I + \sigma_3 \big]$ and $\half I$ respectively.
However, the image of the Bloch sphere is now the
ellipsoid
\be
  \half x^2 + \half y^2 + (z-\half)^2 = \frth
\ee
shown in Fig.~\ref{fig:amp}.
Schumacher and Westmoreland \cite{SW2} studied this channel and
observed that its maximum capacity of $0.4717$ is attained along
a horizontal line whose image is at height $z = 0.596$.

 \smallskip

\begin{figure}
\epsfxsize=9cm
\epsfbox{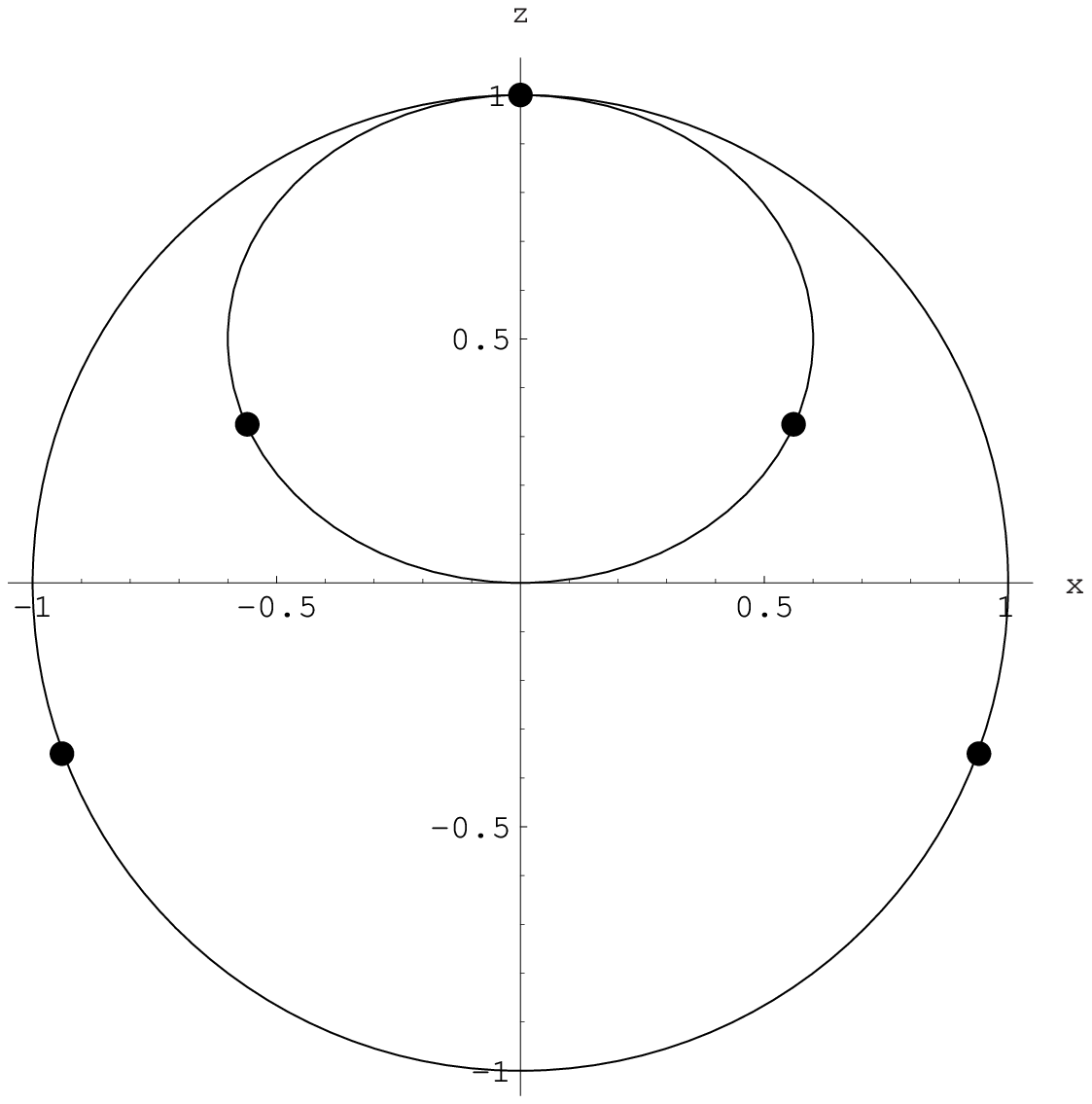}
\caption{The x-z plane of the stretched channel,
with $\mu = 0.5$, showing the optimal inputs
and their images.}
\label{fig:str}
\end{figure}

These results suggest that we begin with a shifted depolarizing
channel (\ref{eq:dep}) and continuously deform it into an amplitude
damping channel  (\ref{eq:amp}) by
stretching the sides to obtain a channel of the form
\be  \label{eq:str.gen}
  \Phi_{\str}(\rho) & = & \half
 \big[I + (s w_1, s w_2, (1-\mu) + \mu w_3) \dtsig \big]
\ee
with $\mu \leq s \leq \sqrt{\mu}$.  Such a channel is shown in
Fig.~\ref{fig:str}.   As $s$ increases from $\mu$ to $\sqrt{\mu}$ there
should be a point at which the capacities restricted to vertical and
horizontal lines  become equal.  Each of these two-state ensembles,
which we denote ${\cal E}_V$ and ${\cal E}_H$ respectively, defines a
unique point on the z-axis, namely the height of the average output
state. If we now take the average of these two ensembles, the resulting
$\chi$-quantity is
\be \label{eq:av.capac}
\lefteqn{ \chi \big[ \half \big( {\cal E}_V + {\cal E}_H \big) \big]
 = \half (C_V + C_H)}~~~~~ \\ & ~ & +
     S \big[ \half \Phi \big( \rho_V + \rho_H \big) \big]
   - \half  S[\Phi( \rho_V)] - \half S[\Phi( \rho_H)].    \nonumber
\ee
Since the entropy is  strictly concave, 
$\chi \big[ \half \big( {\cal E}_V + {\cal E}_H \big) \big]$
will be strictly greater than  $\half(C_V + C_H) = C_V = C_H$ 
{\em unless} 
$\Phi_( \rho_V) = \Phi( \rho_H)$.  We have verified that
$\Phi_{\str}( \rho_V) \neq \Phi_{\str}( \rho_H)$   for the crossings
associated with a range of values of $\mu$, including $\mu = 0.5$.  
Since the average ensemble $\half \big( {\cal E}_V + {\cal E}_H \big)$
contains four states, this implies that
the 4-state capacity $C_4$ is strictly greater than the restricted
two-state capacity $C_V = C_H$.  However, the four input states
used here all lie in the same plane, in fact on a circle of the
Bloch sphere.   The corresponding circle of density matrices
lies within a 3-dimensional subspace.   By a straightforward
adaptation of the general arguments in  \cite{Davies},
the maximum capacity on a circle is achieved with at most three
states.   Numerical studies of examples of this type have
confirmed that the 3-state capacity is strictly greater than the
(unrestricted) 2-state capacity, and that $C_2 < C_4 = C_3$.
Note that even when the 4-state expression for $\chi$ is always
strictly less than  $C_3$, the restricted 3-state and 4-state
capacities, $C_3$ and $C_4$ respectively, are equal since one
can always add a fourth state with arbitrarily small
probability  to the optimal 3-state ensemble.

\begin{figure}
\epsfxsize=9cm
\epsfbox{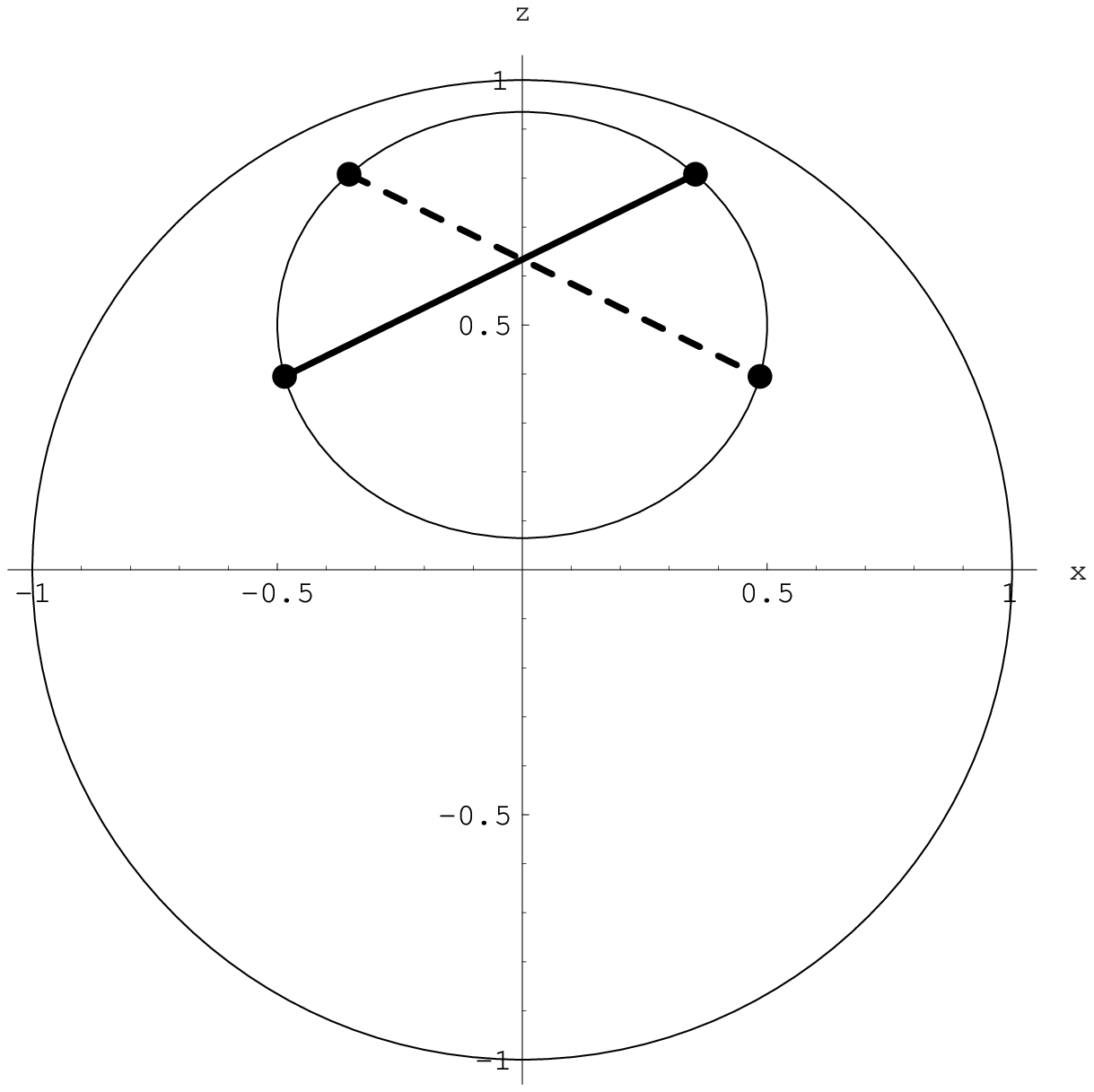}
\caption{The x-z plane of the squeezed channel with
$\mu = 0.5$, $q = 0.435$
showing an ensemble on a skew line and its reflection.
If $\chi$ is not optimized at the crossing,
 averaging to obtain a 4-state ensemble will increase $\chi$}
\label{fig:sqz}
\end{figure}

To complete the argument above, we need to show that
the 2-state capacity $C_2$ actually equals the restricted capacity
$C_V = C_H$. Suppose instead that there is an ensemble on a skew line for
which the $\chi$-quantity (\ref{eq:holv}) exceeds
$C_V = C_H$. Then by symmetry, there will be another ensemble
(obtained by rotating $180^{0}$ around the z-axis)
with the same $\chi$.  Unless
their lines cross along the z-axis, and unless
the average ensemble occurs at exactly this crossing, one will
again be able to average the two ensembles (as shown in
Fig.~\ref{fig:sqz}) to obtain a  4-state $\chi$ value which is
strictly larger. This argument shows that, unless there is a special
degeneracy,  the 4-state capacity always exceeds the
2-state capacity for channels of this type, i.e., those with equal
vertical and horizontal capacity.

 \smallskip

The stretched  channel (\ref{eq:str.gen}) was studied numerically
for $\mu = 0.5$.  When $s = 0.6015$,
$C_V( \Phi_{\str}) = C_H( \Phi_{\str}) =  0.32193$ with average
output states at  $z = 0.6$ and $z = 0.58$ respectively.
Thus, $\Phi_{\str}(\rho_V) \neq \Phi_{\str}(\rho_H) $, implying 
that $C_4 > C_2$  by the argument above.

Detailed numerical studies were then carried
out for $s = 0.6$, yielding
$C_2(\Phi_{\str}) = C_V(\Phi_{\str}) =  0.32193.$
The maximum capacity $C(\Phi_{\str}) =  0.32499$ is achieved
with a three state ensemble consisting of the input state $(0,0,1)$ with
probability $p = 0.4023$, and two symmetric states 
$(\pm .93681, 0, -.34984)$
each with $p= .29885$  (where we denote states by 
vectors in ${\bf R}^3$ via the correspondence
$\rho = \half[I + \bw \dtsig]$.  The optimal inputs and their images
are shown in Fig.~\ref{fig:str} .) 
Although the difference
$C_3(\Phi_{\str}) - C_2(\Phi_{\str}) = 0.003$ is small,
a three state ensemble is definitely needed to achieve the maximum.
The Shannon capacity was also computed
and shown to be achieved with the same two-state ensemble that
gives $C_V$.


Some insight, and a mechanism for constructing additional
examples, can be obtained by considering
the so-called quantum-classical (QC) and classical-quantum (CQ)
channels introduced by Holevo \cite{Hv3}.  For
qubits, both QC and CQ channels contract the
Bloch sphere to a line.  Up to a rotation, a QC channel has the form
\bee
   \Phi_{QC}(\rho) = \half \big[I + \big( t_3 + \mu \, w_3 \big) \dtsig
\big]
\eee
with $|t_3| + |\mu| \leq 1$, and its image states are
classical in the sense that they are diagonal.
When  $|t_3| = 1- |\mu|$,
 the line is translated parallel to itself until it touches
the Bloch sphere. In the examples considered above, the vertical
capacity is equivalent to that of a QC channel that is extreme
in the sense of reaching the Bloch sphere.

Similarly, up to a rotation, a CQ channel has the form
\bee
   \Phi_{CQ}(\rho) = \half \big[I + (t_1, t_2, t_3 + \mu \, w_3 ) \dtsig
      \big]
\eee
with $|t_1|^2 + |t_2|^2 +(|t_3| + |\mu|)^2 \leq 1$.
When $t_1$ or $t_2 \neq 0$, states on the image line do not
commute and  retain quantum features.
Indeed, channels of the form
$ \half \big[I + \sqrt{1 - \mu^2} \, \sigma_1 + \mu \, w_3 \, \sigma_3
\big]$ were the first \cite{Hv1,Hv3} for which the Holevo
capacity was shown to strictly exceed the Shannon capacity. The horizontal
capacity we have considered is essentially the capacity of a
non-extreme CQ channel whose image lies on a line of the form
$ \half \big[I + \nu w_1 \sigma_1 + t_3 \sigma_3 \big]$
with $|\nu| = 2 s \sqrt{t_3(1-t_3)}$ and $s$ is as in
(\ref{eq:str.gen}).

 \smallskip

Additional examples of channels which require three inputs to
maximize capacity can be constructed by  deforming
shifted depolarizing channels  (\ref{eq:dep}) in various ways.
One can effectively stretch
$\Phi_D$ by taking a convex combination with the
amplitude damping channel $\Phi_{\amp}$ to obtain a channel
 of the form (\ref{eq:str.gen}).
This channel can also be written
as the convex combination of an amplitude damping channel with a
suitable QC channel.  Furthermore,
the channel (\ref{eq:dep}) can be effectively
squeezed down by taking a convex combination of $\Phi_D$
with a suitable CQ channel to obtain a channel of the form
\be  \label{eq:sqz.gen}
 \Phi_{\sqz}(\rho) = \half \big[I + (\mu w_1, q w_2, (1-\mu) + q w_3)
   \dtsig \big] .
\ee
This channel is shown in  Fig.~\ref{fig:sqz}  for $\mu = 0.5, q
=0.435$.

 \smallskip

The behavior observed above for the stretched channel
(\ref{eq:str.gen}) with  $\mu = 0.5$ is generic over a range
of values of $\mu$, i.e., there
is a value of $s$ for which $C_V = C_H$, and a small interval
for which $C_3 > C_2$.   This was confirmed by additional
numerical studies for $\mu = 0.8$.  When $s = 0.84$, the capacity
of $C = 0.62088$ is achieved with the three state ensemble of
$(0,0,1)$ with $p = 0.34415$, and the two equiprobable states
$(\pm 0.942895, 0, -0.333091)$. The two state capacity $C_{2} = 0.61823$
is achieved with the orthogonal inputs $(0,0,\pm 1)$.


Similarly, for a squeezed channel of the form (\ref{eq:sqz.gen}), there
should be a value of $q$ for which $C_V = C_H$ and a small interval
for which $C_3 > C_2$. Numerical studies were again
conducted with $\mu = 0.5 $.  When $q =0.43535$,
$C_V( \Phi_{\sqz}) = C_H(\Phi_{\sqz})  =0.21325$,
suggesting that $C_4 > C_2$.
This channel was studied in detail for $q =0.435$ with results
similar to those for the stretched channel,
but smaller effect. The capacity of $C = 0.2140$ is achieved
with the three state ensemble $(0,0,1)$ with
$p = 0.3310$ and  two symmetric states $(\pm .9534, 0, -.3017)$
each with $p= 0.3345$.   The two state capacity of
$C_2 = 0.2132$ is achieved with a non-orthogonal pair of horizontal
inputs and is therefore greater than the Shannon capacity of $C_{\shan} =
0.2128$ which is again achieved with a pair of orthogonal inputs. The
difference $C_3(\Phi_{\sqz}) - C_2(\Phi_{\sqz}) = 0.0008$,
although  small, is greater than the difference
$ C_2(\Phi_{\sqz}) - C_{\shan}(\Phi_{\sqz}) = 0.0004$ by
a factor of two.

 \smallskip

In general, to optimize (\ref{eq:holv})  one seeks ensembles whose outputs
are near the Bloch sphere, where the second term is small, but whose
average is near the origin where the first term achieves its
maximum of $1$.  For unital maps, both requirements
are compatible and  achieved
with a two-state ensemble whose image lies along a major axis of
the ellipsoid.  Symmetry may permit, but can never require, ensembles
with more than two inputs.   Moreover, any deformation
of a sphere which yields a unique major axis, also gives a unique
optimal ensemble.   For non-unital maps, the generic situation is
still a two-state ensemble, whose image is on a line segment
determined by the competing terms  (\ref{eq:holv}).
 More inputs are required when,
as in the examples above, two quite differently oriented line segments
corresponding to QC and/or CQ channels emerge with nearly equal two-state
capacities.

We expect to also find qubit channels which require four states to
maximize the Holevo capacity.    However,  constructing an example by
further deformations of the channels above  will require a carefully
balanced asymmetric shifting and squeezing.  Either stretching or
shifting alone appears to break the balance and return to a two-state
channel.  Moreover, care must be taken to ensure that the deformed
ellipsoid is one whose parameters satisfy the complete positivity
conditions, as discussed in \cite{RSW}.   Work in this direction
is underway.

 \smallskip

There is another expression for the Holevo capacity,
discovered independently by  Ohya, Petz and Wanatabe \cite{OPW}
and  Schumacher and Westmoreland \cite{SW1,SW2}, in terms
of the relative entropy $H(P,Q) = \tr P \log P - \tr P \log Q$.
\be
 C(\Phi) = \ts{\sup_\gamma}  \,  H[ \Phi(\gamma), \Phi(\rho^*) ]
\ee
where $\rho^* = \sum_j \pi_j \rho_j$  is the average state
  for the optimal ensemble.
 As an immediate corollary, one finds that
the output states in the optimal ensemble are equi-distant from the
output average in the sense of relative entropy, i.e., for the
optimal ensemble $ H[ \Phi(\rho_i), \Phi(\rho^*) ] =  C(\Phi)$ for all
$\rho_i$.   Although the relative entropy is {\em not} a true
metric, it can be a useful measure of the ``distance'' between
two states.   Because determining the states which achieve capacity can be
numerically delicate, the equidistance property of optimal
ensembles provides a useful criterion for 
verifying our numerical work.

Moreover, one expects the symmetry properties of the
channel to be reflected in the optimal ensemble.  For the examples
considered above, if one assumes that the optimal ensemble will be a
triple in a given plane with the North pole as $\rho_1$, then the
ensemble is completely determined by the condition that $\rho^*$ is
on the z-axis and   $ H[ \Phi(\rho_1), \Phi(\rho^*) ]
 = H[ \Phi(\rho_2), \Phi(\rho^*) ] = H[ \Phi(\rho_3), \Phi(\rho^*) ]$.
For channels of the form (\ref{eq:sqz.gen}) the condition
$\mu > q$ implies that the optimal ensemble will lie in the $x$-$z$
plane.  Channels of the form (\ref{eq:str.gen}) are symmetric with
respect to rotation around the $z$-axis, which implies that it
suffices to restrict attention to the $x$-$z$ plane; however, any plane
containing the $z$-axis would also suffice.  In addition, when 
$C_3 > C_2$ the capacity can also be achieved (not just approached) 
with a 4-state ensemble.  For example, when $\mu = 0.8$, the two symmetric
states can be replaced by the three states 
     $(\pm \frac{\sqrt{3}}{2} 0.9534, \half 0.9534, -0.3017) $, 
 $(0,-0.9534, -0.3017)$, each with probability $0.223$.

 \smallskip

To summarize, it has been conjectured that the Shannon capacity of any
qubit channel
is achieved when the inputs are strings formed from products of two
orthogonal pure states.   By contrast, as Fuchs \cite{Fuchs} showed, there
are  channels whose Holevo capacity is achieved with two non-orthogonal
input states. This means that the capacity of such a channel
is enhanced when inputs are strings formed from products of two
non-orthogonal states $\{\rho_0, \rho_1\}$, provided entangled
measurements can be used on the outputs.   We have established
that there are situations in which entangled measurements can
enhance the capacity even further when the products are chosen
from {\it three} non-orthogonal states $\{\rho_0, \rho_1, \rho_2\}$
in a two-dimensional Hilbert space.    Moreover, the effect arises
from a competition between a QC capacity with an asymmetric
probability distribution, and a CQ capacity with a 50/50
distribution.  This suggests, in particular, that the
capacity as a function of all possible inputs has a large flat
region near its maximum.  It may be possible to exploit this
flatness and use a
variety of alphabet distributions with only a small sacrifice in
capacity.


\bigskip

\noindent{\bf Acknowledgment}  The work of C. King
was partially supported by National Science
        Foundation Grant DMS-0101205.
The work of M. Nathanson and M.B. Ruskai was partially supported  by
 the National Security Agency  and
 Advanced Research and Development Activity  under
Army Research Office contract number
   DAAG55-98-1-0374 and by the National Science
        Foundation under Grant number DMS-0074566.

\bigskip

\bigskip

\begin{thebibliography}{~~}

\bibitem{Davies} E.B. Davies, ``Information and quantum measurement"
{\em IEEE Trans.} {\bf TI--24}, 596--599 (1978).

\bibitem{Fuchs} C. Fuchs, ``Nonorthogonal quantum states maximize classical
information capacity" {\em Phys. Rev. Lett}, vol. 79,
pp. 1162--1165, 1997.

\bibitem{Hv1} A. S. Holevo, ``On the capacity of quantum communication
channel", {\em Probl. Peredachi Inform.}, {\bf 15}, no. 4, 3-11 (1979)
(English translation: {\em Problems of Information Transm.}, {\bf 15},
no. 4, 247-253 (1979)).

\bibitem{Hv2} A.S. Holevo,
 ``The capacity of quantum channel with general signal states"
{\em IEEE Trans.  Info. Theory} {\bf 44}, 269-273 (1998).
 preprint (lanl: quant-ph/9611023)

\bibitem{Hv3} A. S. Holevo, "Quantum coding theorems",
{\em Russian Math. Surveys}, vol. 53:6, pp. 1295-1331, 1999.


\bibitem{KR1} C. King and M.B. Ruskai,
``Minimal Entropy of States Emerging from Noisy Quantum Channels"
{\em IEEE Trans.  Info. Theory} {\bf 47}, 192--209 (2001).

\bibitem{KR2} C. King and M.B. Ruskai,
``Capacity of Quantum Channels Using Product Measurements''
{\em J. Math. Phys.} {\bf 42}, 87--98 (2001).

\bibitem{OPW} M. Ohya, D. Petz and N. Watanabe, ``On capacities of quantum
channels" {\em Prob. Math. Stats.} {\bf 17}, 170--196 (1997).

\bibitem{RSW} M.B. Ruskai, S. Szarek and W. Werner,
``An analysis of completely positive trace-preserving maps on
${\cal M}_{2}$" to appear in {\em Lin. Alg. Appl.}.
preprint lanl:quant-ph/0101003.

\bibitem{SW1}   B. Schumacher and M. D. Westmoreland,
``Sending classical information via noisy quantum channels"
{\em Phys. Rev. A} {\bf 56}, 131--138 (1997).

\bibitem{SW2} B. Schumacher and M. D. Westmoreland,
``Optimal Signal Ensembles" preprint lanl:quant-ph/9912122.


\end{thebibliography}
\end{document}